\def\PRL{{ Phys. Rev. Lett.\ }\/}
\def\PRB{{ Phys. Rev. B\ }\/}

\def\etal{{\it et al.~}\/}

\def\be{\begin {equation}}
\def\ee{\end {equation}}
\def\ber{\begin {eqnarray}}
\def\eer{\end {eqnarray}}
\def\bers{\begin {eqnarray*}}
\def\eers{\end {eqnarray*}}
\def\eq{\ =\ }

\makeatletter

\newcommand{\Rmnum}[1]{\expandafter\@slowromancap\romannumeral #1@}
\makeatother

\makeatletter
\newcommand*\env@matrix[1][*\c@MaxMatrixCols c]{%
  \hskip -\arraycolsep
  \let\@ifnextchar\new@ifnextchar
  \array{#1}}
\makeatother

\documentclass[aps,prb,showpacs,superscriptaddress,a4paper,twocolumn]{revtex4-1}
\usepackage{amsmath}
\usepackage{float}
\usepackage{color}

\usepackage[normalem]{ulem}
\usepackage{soul}
\usepackage{amssymb}

\usepackage{amsfonts}
\usepackage{amssymb}
\usepackage{graphicx}
\begin {document}

\title{Type-II Dirac states in full Heusler compounds XInPd$_2$ (X = Ti, Zr and Hf)}

\author{Chiranjit Mondal}
\thanks{These two authors have contributed equally to this work}
\affiliation{Discipline of Metallurgy Engineering and Materials Science, IIT Indore, Simrol, Indore 453552, India}

\author{Chanchal K. Barman}
\thanks{These two authors have contributed equally to this work}
\affiliation{Department of Physics, Indian Institute of Technology, Bombay, Powai, Mumbai 400076, India}

\author{Biswarup Pathak}
\email{biswarup@iiti.ac.in }
\affiliation{Discipline of Metallurgy Engineering and Materials Science, IIT Indore, Simrol, Indore 453552, India}
\affiliation{Discipline of Chemistry, School of Basic Sciences, IIT Indore, Simrol, Indore 453552, India}

\author{Aftab Alam}
\email{aftab@iitb.ac.in}
\affiliation{Department of Physics, Indian Institute of Technology, Bombay, Powai, Mumbai 400076, India}

\date{\today}

\begin{abstract}

We predict three full Heusler compounds XInPd$_2$ (X = Zr, Hf and Ti) to be potential candidates for type-II Dirac semimetals. The crystal symmetry of these compounds have appropriate chemical environment with a unique interplay of inversion, time reversal and mirror symmetry. These symmetries help to give six pairs of type-II Dirac nodes on the C$_4$ rotation axis, closely located at/near the Fermi level. Using first principle calculations, symmetry arguments and crystal field splitting analysis, we illustrate the occurrence of such Dirac nodes in these compounds. Bulk Fermi surfaces have been studied to understand the Lorentz symmetry breaking and Lifshitz transition (LT) of Fermi surfaces. Bulk nodes are projected on the (001) and (111) surfaces which form the surface Fermi arcs, that can further be detected by probes such as angle resolved photo-emission and scanning tunneling spectroscopy. By analyzing the evolution of arcs with changing chemical potential, we prove the fragile nature and the absence of topological protection of the Dirac arcs. Our predicted compounds overcome the limitations of the previously reported PtTe$_2$ class of compounds. 
\end{abstract}


\maketitle

\par {\it  Introduction:}
Collective excitation of elementary electrons warrants several distinct Fermionic behavior in condense matter systems. These Fermions can be classified as Dirac, Majorana, Weyl, triple-point, nodal-line semimetal etc. Although the Dirac Fermions have been realized in high energy experiments, the Weyl and Majorana had not so far been observed until the discovery of topological semimetals and superconductors in low energy condense matter systems. Dirac semimetals (DSM)\cite{intro1,intro2} and Weyl semimetals (WSM)\cite{intro4,intro5} are the well known topological materials with surface states (SSs) driven Fermi arcs (FAs). In DSM (WSM), four (two) bands cross each other near the Fermi level (E$_F$). Such band crossing introduce massless Fermion nature in the low energy excitation which brings several spectacular physical properties, such as, close and open FAs,\cite{FA2018} spin Hall effect (SHE),\cite{qshe2011} chiral anomaly\cite{chiral2016,chiral2018} etc. If two bands of a given crystalline material having inversion symmetries (IS) and time-reversal symmetries (TRS) cross each other in the momentum space, it can host a Dirac like quasi particle excitation near the nodal point. Such quasi particle excitation can be well captured by the relativistic Dirac equation, hence the name DSM. Whether the DSM phase is stable or fragile, it depends on crystal symmetry. For instance, under the TRS preserving condition, the Dirac nodes are stable in systems with a particular space group symmetry. The four-fold Dirac node splits into two two-fold Weyl nodes with opposite chiralities if either of the IS or TRS is broken.

\begin{figure}[t]
\centering
\includegraphics[width=\linewidth] {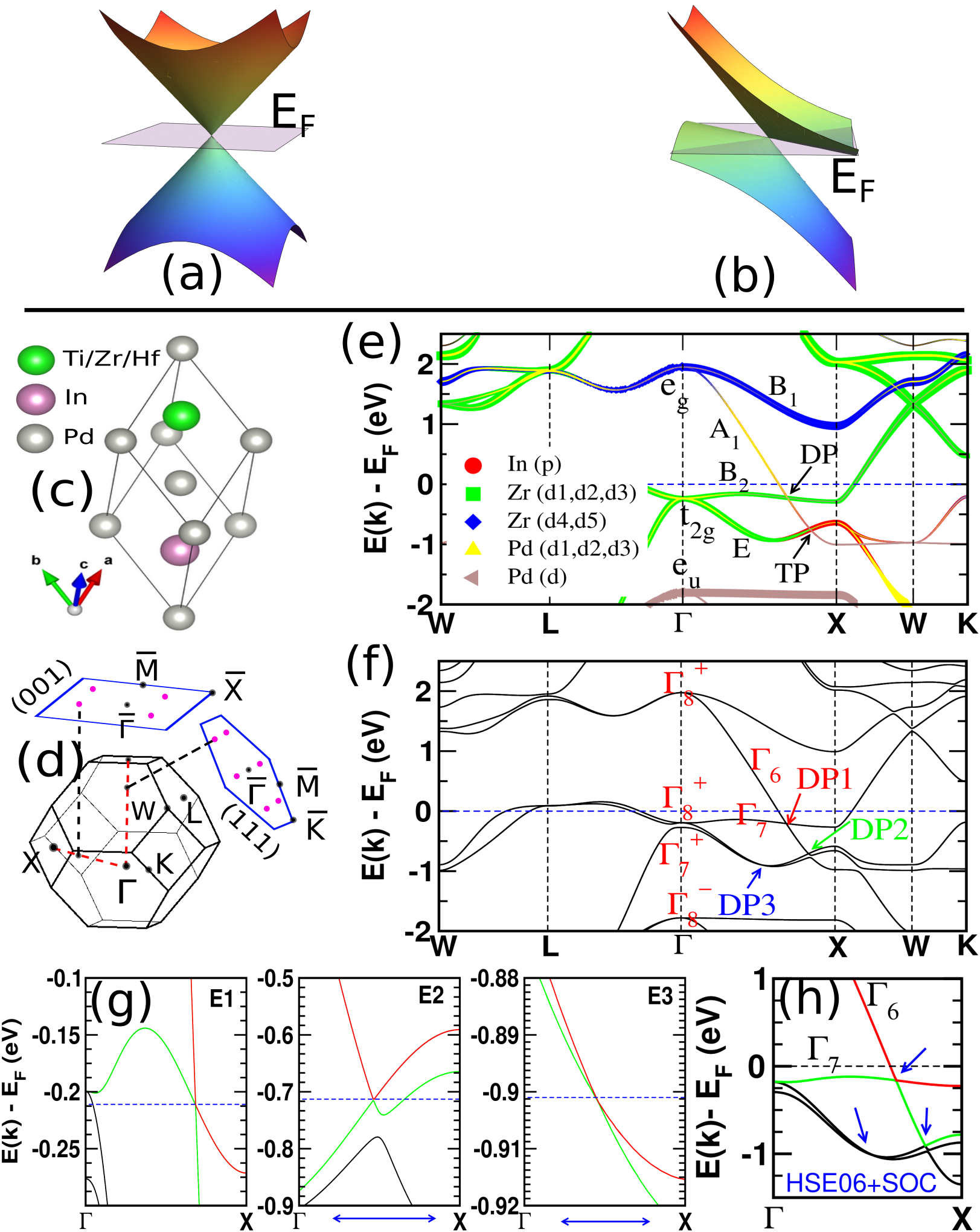}
\caption{(Color online) (a) Type-I Dirac node  and corresponding point like Fermi surface (FS) at the nodal point. (b) Type-II  Dirac node and corresponding contour like FS. (c) Crystal structure of XInPd$_2$. (d) Bulk and surface Brillouin zones. For ZrInPd$_2$, (e) orbital projected bulk band structure without SOC. (f) Bulk band structure with SOC. (g) Zoomed view of (f) near the three nodal points DP1, DP2, and DP3 at energy E1, E2 \& E3 respectively. (h) HSE06+SOC band structure along $\Gamma$-X direction.}
\label{fig1}
\end{figure}

So far, DSM or WSM compounds with point like Fermi surface (FS) (Fig.~\ref{fig1}(a)) at the nodal points have been studied extensively and referred as type-I semimetals. \cite{typeI-1, typeI-2, typeI-3, typeI-4} There are classes where these Dirac or Weyl nodes get sufficiently tilted in the momentum space shifting the point like FS to a contour like FS (Fig.~\ref{fig1}(b)). Such contour like FS yields strikingly different physics as compared to type-I semimetal, and are called type-II DSM or WSM. Details of a general model Hamiltonian describing the band topology for type-II DSM and WSM are given in Sec. I of the supplemental material (SM).\cite{supp}

{\par}Till date, several type-I DSM compounds have been theoretically predicted and experimentally verified through the photo-emission and transport measurements. In contrast, only a handful number of type-II DSM\cite{typeII-1,typeII-PtTe2,typeII-PtSe2,typeII-ATe2} compounds are investigated. The major problem of the existing compounds are the position of nodal points which are far from the E$_F$ and the presence of additional trivial Fermi pockets. For example, extensively studied compounds PtSe$_2$\cite{typeII-PtSe2} and PtTe$_2$\cite{typeII-PtTe2, PtTe22017} have type-II Dirac node around 1 eV below E$_F$ with several trivial band crossings. VAl$_3$\cite{VAl3} is another type-II DSM class where nodal points lie above E$_F$ and restrict the photo-emission experiments to probe them. In this paper, using \textit{first principle calculations}, (see Sec. II of SM \cite{supp} for computational details) we predict three full Heusler alloys XInPd$_2$ (X=Ti, Zr, Hf) to showcase the type II DSM properties. This study can guide not only the future photoemission experiments to probe the SSs, but also shed light on the currently debated topic of fragile nature of Fermi arcs and their associated topological origin. Although there exists another compound in this class YPd$_2$Sn,\cite{YPd2Sn} but a detailed study of topological Fermi arcs, SSs and bulk FS driven Lifshitz  transition (LT) is lacking. One of the important features of ZrInPd$_2$ and HfInPd$_2$ is that they show superconducting phase transition at temperature 2.19 K and 2.86 K respectively.\cite{SC2009XInPd2,SC2012XInPd2} Previously, topological superconductivity (TSC) have been studied in a general framework of Fermi loop (FL) topology and C$_n$ rotational symmetry lowering in DSM. \cite{TCS-1, TCS-2} For example, Dirac compound Cd$_3$As$_2$ shows superconductivity under pressure.\cite{Cd3As2-SC} With the unique orbit-momentum locking near the nodes and the C$_4$ to C$_2$ rotational symmetry lowering creates a gap at the nodal points which is speculated to stabilize the TSC phase by increasing the condensation energy.\cite{TCS-1} In contrast to the type-I DSM, bulk FS of type-II DSM is composed of non-trivial electron and hole pockets which may contribute to the formation of the Cooper pairs and allow the compound to become superconducting.\cite{LT2017}  Although, very recently, PdTe$_2$\cite{PtTe22017} and YIn$_3$\cite{YIn3} type-II DSM class have been put forward to ignite the TSC study but these two classes face certain limitations similar to those of PtTe$_2$ and PtSe$_2$, as discussed above. Our full Heusler compound ZrInPd$_2$ is much more superior in the above contexts as it's type-II Dirac node lie almost at E$_F$ with relatively less number of Fermi pockets.


{\par} {\it Results and Discussions:} The full Heusler compounds XInPd$_2$ belong to the space group Fm$\overline{3}$m\cite{FHABook} where X and In have the equivalent Wyckoff positions (0,0,0) and (1/2,1/2,1/2) to form a rock-salt structure and Pd takes sits at (1/4,1/4,1/4) and (3/4,3/4,3/4). The primitive unit cell is shown in Fig.~\ref{fig1}(c). The bulk Brillouin zone (BZ) and projected (001) and (111) surface BZ are shown in Fig.~\ref{fig1}(d).

\begin{figure}[t]
\centering
\includegraphics[width=\linewidth]{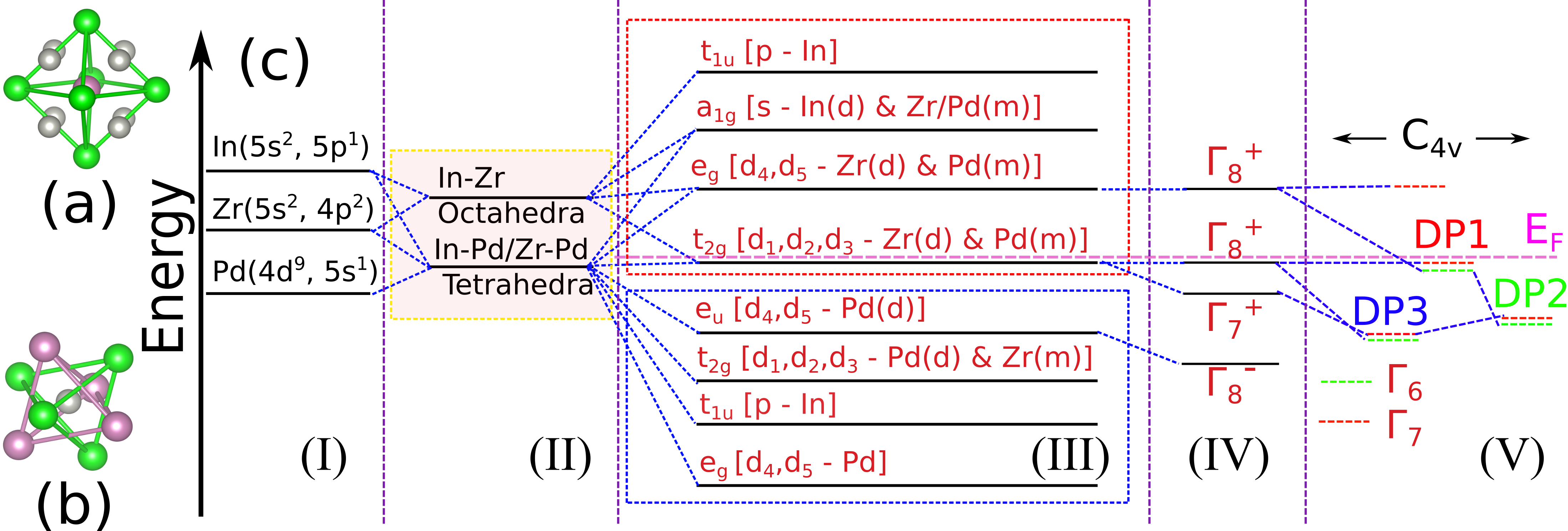}
\caption{(Color online) (a) Zr-In Octahedra  (b) Two inter-penetrating tetrahedral configuration in ZrIndPd$_2$. (c) Effect of orbital hybridization, crystal field splitting and formation of Dirac nodes. Region (I) is the atomic energy levels according to Aufbau principle, region (II) represents the energy level formation of octahedra and tetrahedra. Region (III)  \& (IV) describe the effect of crystal field splitting at $\Gamma$ point without and with SOC respectively. The bands are designated with their irreducible representations (IRs). Region (V) shows the band representations and formation of nodal points along $\Gamma$-X direction or C$_{4v}$ axes. `d' and `m' represent Major and minor contributions respectively from a particular atom.}
\label{fig2}
\end{figure}

The general formation mechanism of Dirac Fermion states have been discussed in Sec. III of SM\cite{supp} considering the point group symmetry of XInPd$_2$. We took ZrInPd$_2$ as a test case and discussed the detail calculated results. Figure~\ref{fig1}(e) shows the band structure of ZrInPd$_2$ without spin orbit coupling (SOC). At the $\Gamma$ point near E$_F$, the major contribution comes from the Zr-t$_{2g}$ orbitals and band above this has dominant Zr-e$_g$ contribution. Zr-t$_{2g}$ split into E and B$_2$ bands along $\Gamma$-X direction whereas e$_g$ transform into two singly degenerate A$_1$ and B$_1$ bands, denoted according to the IRs of C$_{4v}$ little group. A$_1$ and B$_2$ bands cross each other to form a 2-fold nodal point and A$_1$ intersect with E to form a 3-fold nodal point as shown in Fig.~\ref{fig1}(e). Inclusion of SOC doubles the eigen space according to double group representation of C$_{4v}$. Therefore, e$_g$ transform as: $e_g \rightarrow \Gamma_{8}^{+}$  and $e_u\rightarrow \Gamma_{8}^-$. However, t$_{2g}$ splits into $\Gamma_{8}^+$ and $\Gamma_{7}^+$ (i.e, $t_{2g}\rightarrow\Gamma_{8}^{+} \oplus \Gamma_{7}^+$,) at $\Gamma$ point. In contrast, along $\Gamma$-X direction, A$_1$ and B$_2$ transform into $\Gamma_6$ and $\Gamma_7$ respectively, whereas E transform as: E$\rightarrow \Gamma_6 \oplus \Gamma_7$ as shown in Fig.~\ref{fig1}(e,f). Note that the 2-fold $\Gamma_{7}^+$ IR in Fig.~\ref{fig2}(f) propagate as $\Gamma_7$ and simultaneously  4-fold $\Gamma_{8}^+$ goes into $\Gamma_6$ and $\Gamma_7$ along $\Gamma$-X direction. The dimensional degeneracy of $\Gamma_6$ and $\Gamma_7$ IRs are two. Hence, accidental band degeneracy of these $\Gamma_6$ and $\Gamma_7$ bands form three Dirac nodes along $\Gamma$-X as denoted by DP1, DP2 and DP3 in Fig.~\ref{fig1}(f). Figure~\ref{fig1}(g) shows the zoomed view of DP1, DP2 and DP3. For DP1 and DP3, the electron (green) and hole (red) bands, which have similar slope, cross each other to form type-II Dirac nodes. However, DP2 is the type-I like Dirac node owing to the opposite electron and hole band slope. The type-II Dirac node (DP1) lies almost at E$_F$ with small Fermi pockets away from the nodal point. The nature of all these nodal points have been explained using the effective \textit{k.p} Hamiltonian, as discussed Sec. IV of SM.\cite{supp} The Chern number (topological index) for a Dirac node is zero as the nodal point can be considered as superposition of two Weyl nodes with opposite topological charges $\pm$1. Vanishing of  Chern number for a Dirac node and its consequences on the Dirac Fermi arcs will be discussed extensively in the subsequent sections. Unlike Na$_3$Bi\cite{A3Bi2012} or Cd$_3$As$_2$,\cite{Cd3As22013} where non-trivial band inversion ({\it Z$_2$}=1) harbors Dirac crossings, our predicted XInPd$_2$ do not show band inversion and hence the Dirac nodes are manifested by accidental trivial band crossings. Further, to make sure about the topology and location of these Dirac nodes, a more accurate HSE06 level calculations are also carried out. This calculations give similar results as that of PBE calculation with Dirac nodes more closer to Fermi level, as shown in Fig.~\ref{fig1}(h).

{\par} To reconfirm the formation of DSM phase in ZrInPd$_2$, we shall now illustrate the above discussion based on the structural and chemical environmental dependent crystal field theory. In full-Heusler alloy, e.g., ZrInPd$_2$, In atom sits at the center of an octahedra formed by Zr atoms (which sit at the six faces of FCC lattice) as shown in Fig.~\ref{fig2}(a). Furthermore, Zr and In atoms form two inter-penetrating tetrahedra (mutually rotated by 90$^{\circ}$) keeping Pd atom at the center of tetrahedra as shown in Fig.~\ref{fig2}(b). The atomic energy level distributions and the effect of crystal field splitting are shown in Fig.~\ref{fig2}(c). The shaded box in region II is the pictorial representation of energy levels of octahedra and tetrahedra in the lattice. Region III represents the effect of orbital hybridization and crystal field splitting in the absence of SOC at the $\Gamma$ point. In region III, the red block (above E$_F$) corresponds to the energy levels, mainly contributed by octahedral symmetry and the blue block (below E$_F$) contain the energy levels corresponding to tetrahedral symmetry. The crystal field splitting of $d$-orbital, due to the O$_h$ environment (in the red block), is further emphasized. The t$_{2g}$ orbitals which are mainly contributed by Zr, lie just below E$_F$ whereas the e$_g$ orbital lie above E$_F$. Here, in region III, d$_1$ d$_2$, d$_3$, d$_4$ and d$_5$ represent the d$_{xy}$, d$_{yz}$, d$_{xz}$, d$_{x^2-y^2}$, and d$_{z^2}$ orbitals respectively. Above the e$_g$ level, a$_{1g}$ and t$_{1u}$ are mainly contributed by In $s$ and $p$ like orbital respectively. This is consistent with the point group formalism because the basis function for a$_{1g}$ is spherically symmetric whereas it is linear  for t$_{1u}$. On the other hand, the lower blue block shows the $d$ orbital splitting in tetrahedral environment. Due to the tetrahedral splitting, the e$_g$ orbital falls below the t$_{2g}$ orbital. Note that, though the tetrahedra does not posses IS but the presence of global IS of the crystal structure enforces the definite parity sates to the energy levels of tetrahedra. Region IV shows band splitting due to SOC. Inclusion of SOC, transform the $t_{2g}$ octahedra level as: $t_{2g}\rightarrow\Gamma_{8}^{+} \oplus \Gamma_{7}^{+}$. On the other hand $e_g$ goes to $\Gamma_{8}^{+}$. Furthermore, the e$_u$  level transform as $\Gamma_{8}^-$. Region V represents the transformation of bands along $\Gamma$-X direction starting from $\Gamma$ point. The detail mechanisms on formation of three Dirac nodes (DP1, DP2 and DP3) have been discussed in previous paragraph.

\begin{figure}[t]
\centering
\includegraphics[scale=0.25]{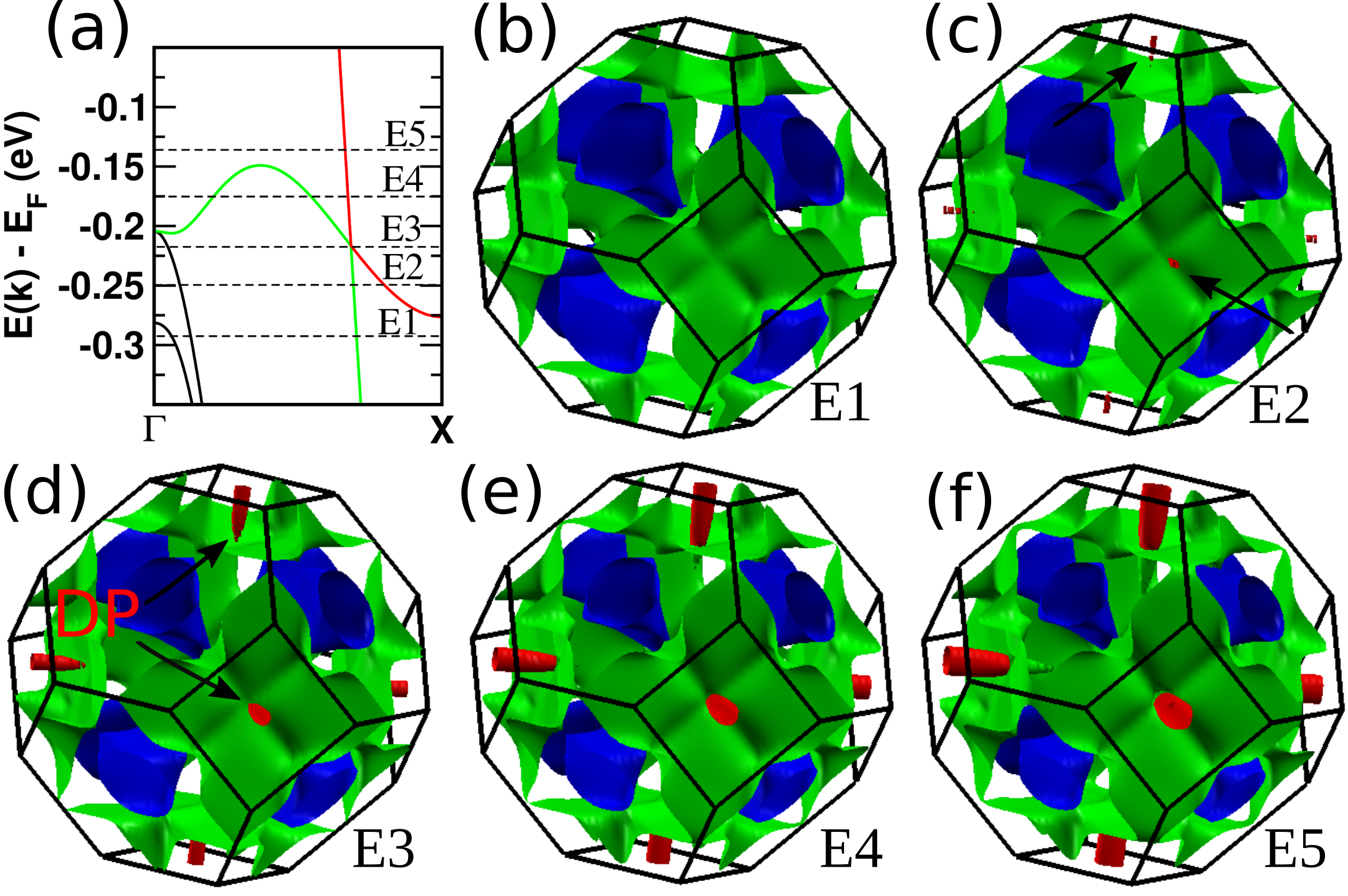}
\caption{(Color online) (a) Zoomed band structure of ZrInPd$_2$ along $\Gamma$-X direction. E1-E5 are different energy cuts around the type-II Dirac node DP1. (b-f) Bulk Fermi surface at different energy cuts E1-E5.}
\label{fig3}
\end{figure}

\begin{figure*}[t]
\centering
\includegraphics[width=\textwidth]{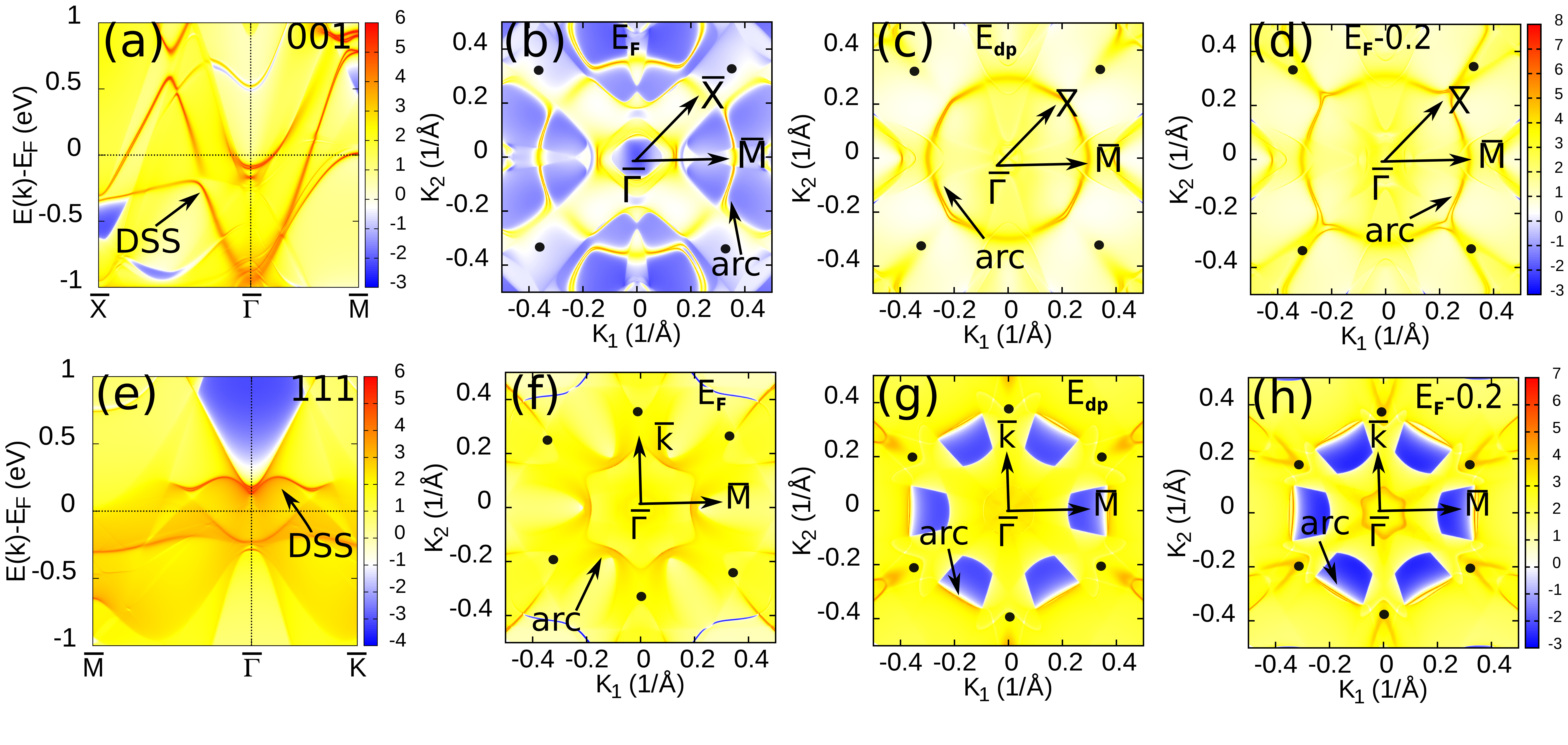}
\caption{(Color online) Surface states (SSs) and Fermi arcs (FAs) of ZrInPd$_2$ using Green's function method. (a) SSs of (001) surface and (b-d) corresponding FAs for different energy cuts. (e) SSs of (111) surface and (f-h) corresponding FAs for different energy cuts.}
\label{fig4}
\end{figure*}
{\par}We will now proceed to understand the bulk FS topology, Lifshitz transition and Lorentz symmetry breaking, for which bulk FS near nodal point DP1 is simulated. The simplest Hamiltonian describing a type-II Weyl node is, \textit{H}$\eq$\textit{c}\textbf{$\vec{\sigma} \cdot \vec{p} $}+\textit{v}\textit{p$_z$}.\cite{LT2017,LT2018} The second term in \textit{H} tilt the cone along z-direction in the momentum space depending on the relative magnitude of \textit{v}. For \textit{v}$\eq$0, the cone is not tilted, it become a type-I semimetal which have point like FS at the nodal point (Fig.~\ref{fig1}(a)). For 0$<$\textit{v}$<$\textit{c}, the cone is tilted. If \textit{v}$>$\textit{c}, the cone is over tilted which results the Fermi level to cross the electron and hole bands, forming contour like Fermi surface connected by the Weyl point, called type-II semimetal (Fig.~\ref{fig1}(b)). The Lifshitz transition occurs for \textit{v}=\textit{c} between these two types of cone. The above scenario is similarly applicable for Dirac nodes but need extra crystalline symmetry to stabilize the Dirac point. Figure~\ref{fig3}(a) shows a zoomed view of bulk band structure along $\Gamma-$X with five energy cuts E1-E5. Figure~\ref{fig3}(b-f) shows the bulk Fermi surface for respective energy cuts. The red(green) bands indicate the electron(hole) type. For E1 energy cut, the electron type FSs are completely absent where as for E2, a tiny electron FS appear as shown by the arrowhead. E3 corresponds to the nodal point energy and the type-II Dirac nodes are indicated by arrowhead in Fig.~\ref{fig3}(d). The electron band contribution in the FSs for E4 and E5 energy cut increases progressively. Further,  FS maps have been projected on a 2D plane to get clear view on the evolution of electron and hole pockets near the DP as shown in Fig. S2.\cite{supp} The FSs evolution justify the tilting of Dirac nodes in the momentum space and breaking of Lorentz symmetry. Here, the crystalline symmetry allows such tilting of cones in low energy system without respecting the Lorentz invariance. Although, DP1 in Fig.~\ref{fig1}(f,h) indicates a type-II Dirac cone, the band effective mass of $\Gamma_{6}$ and $\Gamma_{7}$ clearly hints that the cone is in near critical region (i.e, \textit{v}=\textit{c}) between type-I and type-II Dirac states. Similar observations of Fermi surface topology have previously been seen in type-II Dirac and Weyl semimetals and commonly referred as Lifshitz transition in Ref. [\cite{YPd2Sn,LT2015}]. Such transition is unique by itself (responsible for exotic phenomenon), and is attributed to the occurrence of superconductivity in ZrInPd$_{2}$.

{\par}Similar to the WSM, DSMs also hold the signature of bulk band degeneracies onto its surfaces. Since one Dirac node can be viewed as superposition of two Weyl nodes, double Fermi arcs (FAs) are expected on the bulk projected surfaces. The topological character of such DSM phase can further be understood from the fragility nature of the surface FAs. Hence surface dispersion and FAs of DSM are worthy of careful investigation both from theoretical and experimental front. To reveal the topological nature of FAs, we investigated the (001) and (111) surfaces. There are six Dirac nodes situated along  the six $\Gamma$-X direction in the first BZ, as shown in the Fig.~\ref{fig1}(d). Of these six, four are projected on the (001) surface (indicated by pink dot along $\overline {\Gamma}-\overline{X}$ direction on (001) surface) and the other two on the $\overline {\Gamma}$ point. If we consider that these projected nodes are the source and sink of the surface arcs (though it is not absolutely true for DSM nodes, that will be discussed in detail in the next paragraph), it is expected to get a square like close Fermi loop on (001) surface. However, the (111) surface contains six projected Dirac points as indicated by pink dot along $\overline {\Gamma}-\overline{K}$ direction on (111) surface BZ. Hence, a hexagonal shaped FA expected on this surface. Figure~\ref{fig4}(a) shows the (001) SS of ZrInPd$_2$ for Zr-In terminated surface. FAs of (001) surface for different energy cuts are shown in Fig.~\ref{fig4}(b-d). Fig.~\ref{fig4}(e) shows the (111) SS for Zr terminated surface and Fig.~\ref{fig4}(f-h) the corresponding FAs at different energy cuts. In Fig.~\ref{fig4}(b-d \& f-h), arcs are simulated using green's function method at E$_F$, energy of DP1 (E$_{dp}$) and 0.2 eV below E$_F$ (E$_F$-0.2) to show their evolution on both (001) and (111) surfaces. The evolution clearly shows the fragile nature of Dirac arcs and further hints about their topological features. We have also calculated the SSs of both (001) and (111) surface using slab method within the {\it ab-initio} framework (see Fig. S1 (a,b) in SM \cite{supp}). Details about the slab model calculation is discussed in Section V of SM.\cite{supp}

{\par}M. Kargarian \etal\cite{Kargariana2016} and Yun Wu \etal\cite{YunWu2019} have proposed the fragile nature of Dirac arcs which deform into a Fermi contour, strikingly different from the concept of Chern number protected Weyl arcs and similar to the surface states of a topological insulator. Such Fermi contour may convert to a loop and can also merge into the bulk projected surface Dirac nodes upon the variation of chemical potentials on the surface.\cite{YunWu2019} The fragile nature of Dirac FAs supports the zero Chern number of Dirac nodes and the absence of topological index mediated protections (although it can be stabilized by certain crystalline symmetry). Figure~\ref{fig4}(b) is the Fermi arc exactly at the E$_F$ on (001) surface. It is clear from the figure that the  arcs are completely disconnected from the bulk projected surface Dirac nodes and they do not even form a close loop. This is different from the nature of Weyl arcs and indeed justify the absence of topological robustness in the bulk system. Interestingly, upon varying the chemical potentials, arcs deformed into the Fermi loop which is again not connected with the projected Dirac nodes, as observed in Fig.~\ref{fig4}(c). The Fermi loop, however, sink into surface DPs with further lowering of chemical potential, as shown in Fig.~\ref{fig4}(d). For (111) surface, the topological nature of arcs, as shown in Fig.~\ref{fig4}(f-h), are similar to that of (001) surface . The arc form a loop shape in Fig.~\ref{fig4}(f). However, they are connected with bulk states in Fig.~\ref{fig4}(g,h) at other energy windows. Such detailed analysis not only guide the experimentalists to correctly probe the SSs but also establish a strong base to the largely debated topic of the topological nature of Dirac arcs. The take home message of the entire discussion is that a little perturbation in the bulk crystal that do not disturb the responsible crystalline symmetry for the Dirac state can deform the surface FAs to a disconnected Fermi contour. However, these arcs may not be completely destroyed because of the presence of crystal mirror invariant planes which can further provide mirror Chern number protection.

In addition to ZrInPd$_2$, we have simulated the band structure of TiInPd$_2$ and HfInPd$_2$ as well and found quite interesting type-II Dirac semi-metal features. The type-II Dirac nodes lie very close to E$_F$ for both the compounds, as shown in Fig.~S3(a,b) in SM.\cite{supp} The SSs and Fermi arcs are expected to be similar to ZrInPd$_2$, as they possess very similar bulk band topology. We have discussed the energy location of the DPs for XInPd$_2$ and compared with previously studied type-II DSMs in Sec. VII of SM.\cite{supp}


{\par} {\it Conclusion:} In summary, we have predicted three full Heusler compounds XInPd$_2$ (X = Ti, Zr and Hf) as potential candidates for type-II DSM. Among them, ZrInPd$_2$ and HfInPd$_2$ has been experimentally synthesized and measured to undergo superconducting transition at 2.19K and 2.86K. The position of the nodal points lie at/near E$_F$ which should facilitate for strong response in the transport measurements. Unlike extensively studied PtTe$_2$ class of compounds, our predicted compounds are more superior for experimental investigation because of the location of nodal point in close vicinity of E$_F$ and relatively less number of trivial Fermi pockets. In the present systems, another type-I and type-II Dirac nodes coexist at a little lower energy along $\Gamma$-X. We have carefully studied the bulk FSs at different energy cuts near type-II node (DP1) to investigate the breaking of Lorentz symmetry, tilting of cone and possible Lifshitz transition (LT) of FS. The bulk FSs indicate that ZrInPd$_2$ lie near the boundary of type-I and type-II semimetal (i.e, at the LT region) which could be an underlying reason for the superconducting transition in these systems. SSs and Fermi arcs are simulated on (001) and (111) surfaces of ZrInPd$_2$ to investigate the bulk boundary correspondence and their topological nature. Our detail analysis of Fermi arcs not only guide the experimentalists to reliably probe the SSs but also shed light on the largely debated topic of the topological nature of Dirac arcs. We conclude that a little perturbation in the bulk crystal, that do not disturb the responsible crystalline symmetry for the Dirac state, can deform the surface Fermi arcs to a disconnected Fermi contour. As such, the Dirac Fermi arcs are fragile in nature and does not have topological protection like Weyl arcs. SSs are also simulated using slab model. Thus, we believe that XInPd$_2$ possibly stands as the most ideal material, yet proposed, to host type-II DSM state. We strongly encourage experimental investigations to reconfirm our findings.

{\par} This work is financially supported by DST SERB (EMR/2015/002057), India. We thank IIT Indore and IIT Bombay for the lab and computing facilities. CKB and CM acknowledge MHRD-India for financial support. AA acknowledges IRCC early carrier research award project, IIT Bombay (RI/0217-10001338-001) to support this research.



\end{document}